\documentclass[preprint,showpacs,amsmath,amssymb]{revtex4}

\usepackage{graphicx}
\usepackage{dcolumn}
\usepackage{bm}

\newcommand{\partN}[3]{\frac{\partial^{#3} {#1}}{\partial {#2}^{#3}}}
\newcommand{\partD}[2]{\frac{\partial{#1}}{\partial{#2}}}
\newcommand{\partopN}[2]{\frac{\partial^{#2}}{\partial {#1}^{#2}}}
\newcommand{\partop}[1]{\frac{\partial}{\partial {#1}}}

\begin{document}

\title{Velocity excitations and impulse responses of strings --- Aspects of continuous and discrete models}

\author{Georg Essl\footnote{Electronic mail: georg@mle.media.mit.edu}}
 \affiliation{%
Media Lab Europe\\
Sugar House Lane\\
Dublin 8, Ireland}

\received{\today}

\begin{abstract}
This paper discusses aspects of the second order hyperbolic partial
differential equation associated with the ideal lossless string under
tension and it's relationship to two discrete models. These models are
finite differencing in the time domain and digital waveguide
models. It is known from the theory of partial differential operators
that in general one has to expect the string to accumulate
displacement as response to impulsive excitations. Discrete models
should be expected to display comparable behavior. As a result it is
shown that impulsive propagations can be interpreted as the difference
of step functions and hence how the impulsive response can be seen as
one case of the general integrating behavior of the string. Impulsive
propagations come about in situations of time-symmetry whereas
step-function occur as a result of time-asymmetry. The difference
between the physical stability of the wave equation, which allows for
unbounded growth in displacement, and computational stability, that
requires bounded growth, is derived.
\end{abstract}

\maketitle

\newpage
\section{Introduction}

Our purpose here is to discuss aspects of the relationship of the
solution of the one-dimensional second order wave equation to two
discrete models thereof. The first discrete model is the digital
waveguide model in one spatial dimension. The second discrete model is
a finite difference model in the time domain. In particular we will
also discuss how this relationship explains different behavior between
the discrete models. This relationship has drawn much attention
recently\cite{BBKS03,Bilbao01,Bilbao03,BS03,EK02,Karjalainen03,KE03,KS03,Smith03}

It is shown that the finite difference model can account for solutions
of the wave equation and that these solutions are physically
meaningful.

This allows for a direct interpretation of recent results by
Karjalainen and Erkut \cite{KE03} from the fundamental solution of the
wave equation. Karjalainen and Erkut gave the restricting conditions
necessary to make finite difference models in the time domain and
digital waveguide models connect.

Regarding the stability behavior of the discrete models, the
continuous stability is discussed and it is shown that physically
stable responses of the string may appear unstable in a
discrete model or signal-processing sense.

The paper is structured as follows. First known derivations of the
solution of the wave equation is given, both classically and via the
theory of fundamental solutions of its partial differential operators.
A discrete comparison of the finite difference \cite{Ames92} and the
digital waveguide model\footnote{For a friendly introductory exposition see
Cook\cite{Cook02}.} follows. Smith's text\cite{Smith03} provides the
authoritative summary in of digital waveguides with respect to the
continuous derivations given earlier. We conclude with implications of
these observations.

\section{Solution of the Wave Equation in One Dimensions}

The results in this section are well-known. They are repeated here to
facilitate arguments in the following sections. Of concern is the
general digital simulation of a string under force. The free ideal
string is well-described by the the $1+1$ dimensional d'Alembertian
operator on a scalar field\footnote{We limit our discussion to this
particular form. A note regarding the relationship of this equation to
pairs of first order equations can be found in Appendix
\ref{sec:hidden}.}:

\begin{align}\label{eq:homo}
\Box y(x,t)\stackrel{\text{\tiny def}}{=}(\partopN{x}{2}-c^2\partopN{t}{2})y(x,t)
\end{align}

An external force leads to the inhomogeneous case of the same equation:

\begin{align}
\Box y(x,t)=f(x,t)
\end{align}

Without loss of generality we assume that $c=1$ for our discussion and
hence one gets the factored form of the d'Alembertian:

\begin{align}
\Box=\partopN{x}{2}-\partopN{t}{2}=\left(\partop{x}+\partop{t}\right)\left(\partop{x}-\partop{t}\right)
\end{align}

The factored form suggests the substitution $\xi=x-t$ and $\eta=x+t$ we obtain the canonical form of the wave equation:

\begin{align}
\frac{\partial^2u}{\partial\xi\partial\eta}=\frac{1}{4}f\left(\frac{\xi+\eta}{2},\frac{\eta-\xi}{2}\right)=\phi(\xi,\eta)
\end{align}

This can be directly integrated, yielding \cite{CH68}

\begin{align}\label{eq:canonsolve}
u=\int_{\eta_0}^\eta\int_{\xi_0}^\xi \phi(u,w)\,du\,dw+h_1(\xi)+h_2(\eta)
\end{align}

where $h_1$, and $h_2$ are ``constants of integration''. However, this
notation hides that either one of these constants has been integrated
over with regards to its parameter. Also these are not uniquely
defined functions from a yet undefined functional space but rather any function
from suitable family of functions \cite{CH68}. The derivation
procedure suggests that $h_1(\cdot)$ and $h_2(\cdot)$ are one and
twice differentiable everywhere in the solution space, but which one
is twice differentiable depends in what order the solution has been
integrated over.


The solution of the homogeneous case (\ref{eq:homo}) with initial
conditions $y(x,0)=f(x)$ and $y_t(x,0)=g(x)$ is well known
\cite{Kreyszig99,SS03} to correspond to d'Alembert's
solution\footnote{It is noteworthy to mention that the nomenclature
for ``d'Alembert's solution'' is not universally agreed upon. Some
sources use it to describe the the general form of the integration of
the homogeneous equation in the absence of initial value or force
(also called Cauchy) data (like \cite{Kreyszig99}) whereas in other
sources, for example\cite{EKS99}, it refers to the solution including
Cauchy (initial, boundary, external force) data. We will use it to denote the solution of the initial value problem.}:

\begin{align}\label{eq:ivsolve}
y(x,t)=\frac{1}{2}\left(f(x+ct)+f(x-ct)\right)+\frac{1}{2c}\int_{x-ct}^{x+ct}&g(s)\,ds
\end{align}

Hence initial displacements travel left and right. Initial velocity,
however smears over a widening domain of influence. 


Alternatively this results can be derived using the theory of partial
differential operators. Writing $L=\Box$ and $X=(x,t)$ one arrives
at the generic partial differential operator equation:

\begin{align}\label{eq:opnot}
Lu(X)=k(X)
\end{align}

Here it is important to note that it is no longer required that
$u(\cdot)$ is in the class of twice differentiable functions
$\mathcal{C}^2$, but rather that $u(\cdot)$ and $k(\cdot)$ are
distributions\footnote{Laurent Schwartz, the originator of the theory
of distributions recently wrote an autobiography \cite{Schwartz01}
that includes a readable account of the historical development of his
discovery. We quote two statements from his book that are of interest
relating to this article. He writes, when describing the prehistory of
his discovery \cite[p. 212]{Schwartz01}: ``One of the most important
precursors of distributions was the electrical engineer Heaviside.''
In a section headed {\em Vibrating strings, harmonic functions} he
writes \cite[p. 218]{Schwartz01}: ``We had learned that the general
solution of this equation is of the form $u(t,x)=f(x+vt)+g(x-vt)$,
where $f$ and $g$ are arbitrary functions of one variable. Naturally,
this presupposes that $f$ and $g$ are $C^2$, so as to be able to
differentiate them. What should one think of a function $u$ which
would be analogous except that $f$ and $g$ would be merely $C^1$
(continuous), or not even continuous? Is it a wave or not? I was
obsessed by this question for some time, then I stopped thinking about
it and relegated the question to a corner of my mind for future
reflection.''} or generalized functions. What this means in detail we
will have to defer to expositions available elsewhere
\cite{Edwards95,Hulshof93,EKS99}. For our purpose it is interest to
note that in the theory of distributions jumps and discontinuity are
gracefully and meaningfully included in the formalism including the
definition derivatives of entities like the Dirac-delta
$\delta(\cdot)$ over the space of suitable functions testing for this
property. In addition it makes continuous convolutions a central
operation to the calculation of continuous solution, which make it's
treatment very similar to the study of discrete models \cite{OS89}
which is regularly used for digital waveguide models\cite{Smith03}.

In our treatment here we will closely follow Hulshof's\cite{Hulshof93}
and Joshi and Wassermann's lecture notes\cite{JW03} and also Edwards'
text\cite{Edwards95} which are more accessible than, for example the
technical survey by Egorov, Komech and Shubin\cite{EKS99}.
The interested reader is pointed to the latter for statements of
necessary theorems as well as proofs or detailed references to
original proofs.

The {\em fundamental solution} of the equation (\ref{eq:opnot}) the
effect of the operator $L$ on the distribution $u$ when it sees as
input a Dirac-delta (as noted earlier this is not a function, but a
distribution). In the digital signal processing literature\cite{OS89}
$u$ is called the impulse response of $L$. In a dynamical sense it is
the response to the inhomogeneous equation where the external force
distribution is a Dirac-delta $\delta$. Of immediate interest are
solution forward in time the discussion is restricted to fundamental
solutions in the positive half-plane with respect to time. This will
be indicated by the superscript $+$ to the symbol $\mathcal{E}$ for
the fundamental solution:

\begin{align}\label{eq:derfund}
L\mathcal{E}^+(X)=\delta(X)
\end{align}

It can be proved that the solution of both the homogeneous equation
with initial value data and the inhomogeneous equation with some
external force distribution can be recovered from the fundamental
solution \cite{EKS99}. This is done, in analogy to the impulse
response convolution in digital signal processing \cite{OS89} by
convolution of the fundamental solution with the force distribution
and the initial value data $u(x,0)=f(\cdot)$ and
$u_t(x,0)=g(\cdot)$. In the literature these are call Cauchy data.

The fundamental solution\cite{Hulshof93,Edwards95} of the
one-dimensional wave equation can be derived to be:

\begin{align}\label{eq:fundsol}
\mathcal{E}^+(x,t)=\frac{1}{2}H(t)\left[H(x+t)-H(x-t)\right]
\end{align}

Here $H(\cdot)$ is the Heaviside distribution, which is the
distributional integral of the Dirac-delta distribution
$\delta(\cdot)$. In conventional functional form the Heaviside step-``function'' can be written as\cite{AS72}:

\begin{align}
H(x)=\begin{cases}
0& x<0,\\
\frac{1}{2}& x=0,\\
1& x>0.
\end{cases}
\end{align}


The interpretation of this equation is indeed important because it
indicates, that the ``system response'' of a wave operator to an input
impulse are not isolated traveling wave pulses but traveling
step-distributions. 

It may be convenient to think of the fundamental solution as the
``distributional continuous impulse response''. This makes sense
because the solution of equation can be recovered by convolution of
the fundamental solution with the Cauchy data. The continuous
convolution has the familiar form \cite{Hulshof93,EKS99}:

\begin{align}
u(x)=(\mathcal{E^+}\ast f)(x)\stackrel{\text{\tiny def}}{=}\int_{-\infty}^\infty\mathcal{E}^+(x-s)f(s)\,ds
\end{align}

where $f(x)$ is Cauchy data in one variable.

If both data and the fundamental solution are in two dimensions (as is
possibly the case for external force distributions), one need to
convolve over both variables.

Specifically it can be shown (see \cite{EKS99,Hulshof93}) that for the
set of Cauchy data $u(x,0)=f(x)$, $u_t(x,0)=g(x)$ and $Lu(x,t)=k(x,t)$
one gets the complete solution:

\begin{align}
u(x,t)=u_f(x,t)+u_g(x,t)+u_k(x,t)
\end{align}

with the convolutions:

\begin{align}
u_f(x,t)&= \mathcal{E}^+_t(\cdot,t)\ast f(x)\label{eq:fundsol1}\\
u_g(x,t)&= \mathcal{E}^+(\cdot,t)\ast g(x)\label{eq:fundsol2}\\
u_k(x,t)&= \mathcal{E}^+ \ast k(x,t)\label{eq:fundsol3}
\end{align}

Performing the convolutions yields the the solution 
which is equivalent to the conventional inhomogeneous initial value
solution of the wave equation \cite{Hulshof93}:

\begin{align}\label{eq:dalembertrepresentation}
\begin{split}
y(x,t)=&\frac{1}{2}\left(f(x+t,0)+f(x-t,0)\right)\\
&+\frac{1}{2}\int_{x-t}^{x+t}g(s,0)\,ds\\
&+\frac{1}{2}\int_0^t\int_{x-(t-\tau)}^{x+(t-\tau)}k(s,\tau)\,ds\,d\tau
\end{split}
\end{align}
 
For proofs of uniqueness see Egorov, Komech and Shubin\cite{EKS99}. In
the absence of external forces this reduces to the initial value
solution (\ref{eq:ivsolve}).

The theory of generalized functions for partial differential operators
explains why equations (\ref{eq:canonsolve}), derived for the forced
case, and (\ref{eq:ivsolve}), derived for the homogeneous initial
value case have similar structure. The inhomogeneous case can in a
generalized sense be made to include the homogeneous initial value
problem (also called Cauchy problem). For example, if we symbolically
write $\tilde{k}(x,t)=\delta(t)k(x,t)$ one sees that the external force
response matches the response to the initial velocity.

\begin{align}
u_g(x,t)=\frac{1}{2}\int_{x-t}^{x+t}g(s,0)\,ds\\
u_{\tilde{k}}(x,t)=\frac{1}{2}\int_{x-t}^{x+t}\tilde{k}(s,0)\,ds
\end{align}

Hence an external force distribution which is impulsive in time
$\tilde{k}(x,\cdot)$ is indistinguishable from the equivalent initial
velocity distribution $g(x)$. Conversely it is noteworthy that initial
values do not prescribe a state of the string alone, but also
prescribe a sudden onset of such state. Hence initial values are not
necessarily a free-field solution of the wave equation, that is the
state of a string in the absence of force. Rather impulsive onset
states act like external forces.

In particular one can write the solution of the wave equation for the inhomogeneous initial value problem in the following simple form \cite{JW03}:

\begin{align}\label{eq:fundcauchy}
u(x,t)=\mathcal{E}^+\ast g(x) + \mathcal{E}_t^+\ast f(x) + \int_0^t \mathcal{E}^+_{t-s}\ast k(x,s)\,ds
\end{align}

Next we note the differentiation of distributions
on step-functions (see also the Appendix \ref{sec:distprop}):

\begin{align}
\langle H^\prime, \phi\rangle=\langle\delta,\phi\rangle
\end{align}

observe, using this relationship, that the derivative of the
fundamental solution (\ref{eq:fundsol}) are two propagating
Dirac-deltas, here again in symbolic notation:

\begin{align}\label{eq:funddiraconly}
\mathcal{E}^+_t=\frac{1}{2}H(t)\left(\delta(x+t)+\delta(x-t)\right)
\end{align}

For the current discussion it remains only to point out that the first
contribution of (\ref{eq:fundcauchy}) look like propagating impulses
under differentiation. 

\begin{figure}
\centering
\includegraphics[width=3.375in]{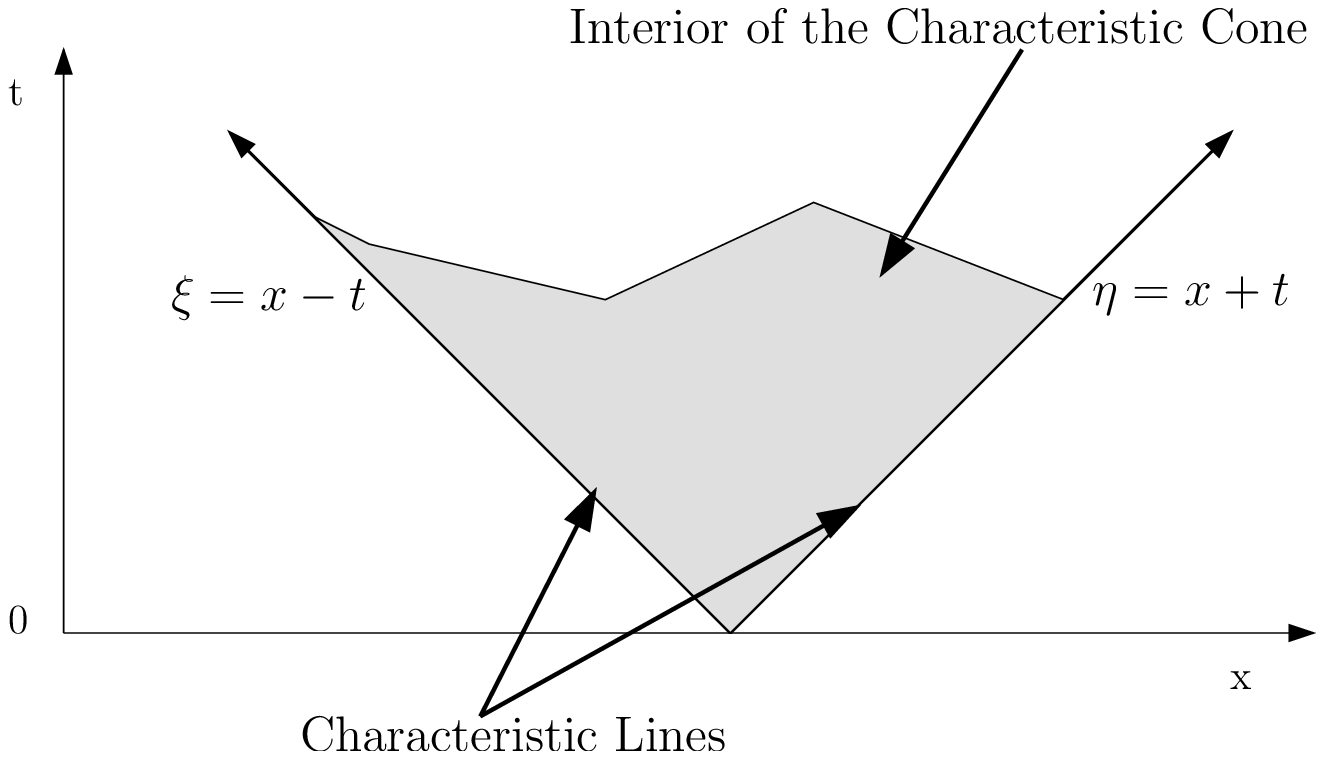}
\caption{The characteristic cone of the one-dimensional wave equation.}
\label{fig:cone}
\end{figure}

From (\ref{eq:fundcauchy}) we see that the solution of the wave
equation will only stay on the characteristic lines $\xi$ and $\eta$ for a
restricted class solutions of the wave equation. Only for very
exceptional cases of initial velocities $g$ and external forces $k$
will the solution not integrate into the domain. Rather generically
one ought to expect them to integrate into the inside of the forward
characteristic cone $x\pm t\geq 0$ as depicted in Figure \ref{fig:cone}.

The condition in which integration into the interior of the
characteristic cone does not occur will be discussed in the discrete
case and in this context has been discovered by Karjalainen and Erkut
\cite{KE03}. For this to be appropriate, care needs to be taken when
assuming the impulse response of the system to have a particular form,
otherwise one ought to expect both contributions at the same
time. Additionally, in a physically realistic situation, when it
cannot be guaranteed that an excitation is of purely
displacement-type, one ought to expect that the dynamics of the system
to integrate and persist over the whole inside of the characteristic
cone. It also is worthwhile to point out that this is the
mathematically consistent solution of the wave equation
\cite{EKS99}. Hence, assuming that the wave equation is a reasonable
model for a given physical situation, one ought to expect such a
behavior to exist and be observable.

As a note, we state, that it is in fact well known that the Huygens'
Principle, the isolated propagation of wave fronts, only holds for
d'Alembertians of odd spatial dimensions greater or equal to $3$
\cite{EKS99,CH68} meaning that only in this case does the fundamental
solution concentrate on the characteristic cone $t^2=|x|^2$. Other
cases, including one spatial dimension\footnote{The case of two
spatial dimensions is relevant, but is not in the main thrust of this
papers discussion. An illustration of the one and two-dimensional wake
of the wave-equation can be found in Graff\cite{Graff91}, p. 220, Fig
4.4.}, have a wave influence inside the characteristic cone. The idea
that the solution would concentrate on the characteristic cone for all
dimensions was held for a long time by mathematicians working on the
wave-equation until Hadamard opened the development and the situation
is since well understood \cite{Garding98}.

Regarding Huygens' principle, a short popular exposition can be found
in \cite{Veselov02} whereas a comprehensive technical exposition can
be found in \cite{Zubelli97}. 

\section{Comparison of Leapfrog and Waveguide Solvers}

Next the leapfrog finite difference scheme \cite{Ames92} will be
compared with the digital waveguide method \cite{Smith03}. The full
treatment of derivations will not be repeated here and the reader is
referred to these sources for details.

Now let us discuss the properties of the so-called leap-frog finite
difference molecule for the wave equation. The explicit time-stepping
equation reads \cite{Ames92,Smith03}:

\begin{align}\label{eq:leapfrog}
\begin{split}
y(n+1,m)=&y(n,m-1)+y(n,m+1)\\&-y(n-1,m)
\end{split}
\end{align}

where the relationship between the discrete time step $T$ and the
spatial discretization $X$ is chosen to satisfy $c=X/T$. In this case
the leapfrog molecule can be shown to be consistent at sampling points
with the wave equation \cite{Ames92}. It can also be shown that
waveguide solutions are solutions of the leapfrog \cite{Smith03}. As
Karjalainen observes, the converse does not hold \cite{Karjalainen03}.

For future discussion we will use the following symbols:

\begin{align}
y_+&=y(n+1,m)\\
y_>&=y(n,m+1)\\
y_<&=y(n,m-1)\\
y_-&=y(n-1,m)
\end{align}


The condition that an impulse at the root of the molecule will only
create responses along the characteristics of the wave can be
expressed by the condition $y_+=0$, i.e. there is no data within the
characteristic domain of the molecule.

Hence the non-integrating molecule condition reads:

\begin{align}
y_<+y_>-y_-=0
\end{align}

From this we get the relationship of waves on the characteristics to their sum:

\begin{align}\label{eq:nonint}
y_<+y_>=y_-
\end{align}

The updating rules for waveguides\footnote{Strictly speaking digital
waveguide synthesis can be formulated in various ways. We will not
use any arguments from transmission-line theory here. For treatment of
those aspects we refer to \cite{Smith03}.} are:

\begin{align}\label{eq:wgupdate}
y_l(n,m)=y_l(n-1,m+1)\\
y_r(n,m)=y_r(n-1,m-1)
\end{align}

with the external force rule:

\begin{align}\label{eq:wgforce}
y_l(n,m)=\frac{1}{2}f(n,m)\\
y_r(n,m)=\frac{1}{2}f(n,m)
\end{align}

The wave reconstruction rule is:

\begin{align}\label{eq:reconwg}
y(n,m)=y_l(n,m)+y_r(n,m)
\end{align}

in response to an external force function $f(n,m)$. If we take an
impulse of height $y_-$ at time $n-1$ we get:

\begin{align}
y_l(n,m-1)=y_l(n-1,m)=f(n-1,m)=\frac{1}{2}y_-\\
y_r(n,m+1)=y_r(n-1,m)=f(n-1,m)=\frac{1}{2}y_-
\end{align}

The reconstructed wave using (\ref{eq:reconwg}) is zero everywhere except at:

\begin{align}
y(n,m-1)=y_l(n,m-1)=\frac{1}{2}y_-\\
y(n,m+1)=y_r(n,m+1)=\frac{1}{2}y_-\\
y(n,m)=y_l(n-1,n)+y_r(n-1,m)=y_-
\end{align}

and we see that the non-integrating case of the leapfrog
(\ref{eq:nonint}) is satisfied with $y_<=1/2 y_-$ and $y_>=1/2 y_-$.

The leapfrog will ``integrate'' whenever condition (\ref{eq:nonint})
is not satisfied. To study the behavior within the characteristic
domain it is first assumed that the elements on the characteristic of the
molecule $y_<$ and $y_>$ are unaltered. That is, the same waves as
before travel outward in the molecule. This leaves us to study an
altered relationship between $y_+$ and $y_-$.

Let $y_-$ be the difference of $y^0_-$, the molecule value for the
non-integrating case (\ref{eq:nonint}), and $\tilde{y}_-$, an assumed contribution to the interior of the characteristic domain. Then we get:

\begin{align}
y_+ &= y_< + y_> - (y^0_- + \tilde{y}_-)\label{eq:h0}\\
 0  &= y_< + y_> - y^0_-\label{eq:h1}
\end{align}

Subtracting (\ref{eq:h1}) from (\ref{eq:h0}) we get:

\begin{align}\label{eq:int}
y_+ &= \tilde{y}_-
\end{align}

Hence the response at at the interior point of the characteristic domain is
constant with regards to the contribution of the incoming wave that violates
the non-integration condition (\ref{eq:nonint}).

To study how this behavior, one can illustrate the response of the leapfrog to an initial $1$:

\begin{equation}\label{eq:intcone}
\begin{matrix}
1 &   & 1 &   & 1 &   & 1 &   & 1\\
  & 1 &   & 1 &   & 1 &   & 1 &  \\
  &   & 1 &   & 1 &   & 1 &   &  \\
  &   &   & 1 &   & 1 &   &   &  \\
  &   &   &   & 1 &   &   &   &  \\
\end{matrix}
\end{equation}

and compare it to an excitation which observes (\ref{eq:nonint}). $y_<=y_>=1$ and $y_-=2$:

\begin{equation}\label{eq:proponly}
\begin{matrix}
1 &   &   &   &   &   &   &   & 1\\
  & 1 &   &   &   &   &   & 1 &  \\
  &   & 1 &   &   &   & 1 &   &  \\
  &   &   & 1 &   & 1 &   &   &  \\
  &   &   &   & 2 &   &   &   &  \\
\end{matrix}
\end{equation}

Observe that (\ref{eq:proponly}) appears to be the sum of traveling
histories and they are time-symmetric around the intersection point. A
time-symmetric solution is an equal contribution to the solution
traveling forward and backward in time and their sum yielding the
complete solution.

D'Alembert's solution (\ref{eq:ivsolve}) can be used to investigate
this observation when writing it in the form following Alpert,
Greengard and Hagstrom\cite{AGH00}:

\begin{align}
\begin{split}
y(x,t)+y(x,-t)=&\frac{1}{2}\left(f(x+ct)+f(x-ct)\right)\\
               &+\frac{1}{2c}\int_{x-ct}^{x+ct}g(s)\,ds\\
               &+\frac{1}{2}\left(f(x-ct)+f(x+ct)\right)\\
               &+\frac{1}{2c}\int_{x+ct}^{x-ct}g(s)\,ds
\end{split}
\end{align}

Taking the time-symmetric sum we get:

\begin{align}
y(x,t)+y(x,-t)=&f(x+ct)+f(x-ct)\label{eq:timesym}
\end{align}

Similarly, by taking the difference, one finds the time-asymmetric case:

\begin{align}\label{eq:timeasym}
y(x,t)-y(x,-t)=\frac{1}{c}\int_{x+ct}^{x-ct}&g(s)\,ds
\end{align}

In the discrete case it is easy to see this property preserved in the
leapfrog case:

\begin{equation}\label{eq:intcone2}
\begin{matrix}
  &   & 1 &   & 1 &   & 1 &   &  \\
  &   &   & 1 &   & 1 &   &   &  \\
  &   &   &   & 1 &   &   &   &  \\
  &   &   & 0 &   & 0 &   &   &  \\
  &   &   &   &-1 &   &   &   &  \\
  &   &   & -1&   &-1 &   &   &  \\
  &   &-1 &   &-1 &   &-1 &   &  \\
\end{matrix}
\end{equation}
\begin{equation}\label{eq:proponly2}
\begin{matrix}
  &   & 1 &   &   &   & 1 &   &  \\
  &   &   & 1 &   & 1 &   &   &  \\
  &   &   &   & 2 &   &   &   &  \\
  &   &   & 1 &   & 1 &   &   &  \\
  &   & 1 &   &   &   & 1 &   &  \\
\end{matrix}
\end{equation}

While (\ref{eq:proponly2}) nicely illustrates the time-symmetry
and the ``interference'' of waves at the interaction point,
(\ref{eq:intcone2}) is insightful as it clearly shows the properties
of a velocity excitation. The displacement vanishes at the interaction
moment, while the temporal slope is maximal. It should be made clear,
that vanishing of data at one time-step in the leap-frog simulation
does not imply velocity solutions. This can be seen if a positive
and a negative impulsive wave cross, creating a time-symmetric situation
that will not integrate inside the domain:

\begin{equation}\label{eq:proponlyasym2}
\begin{matrix}
  &   & -1 &   &   &   & 1 &   &  \\
  &   &   & -1 &   & 1 &   &   &  \\
  &   &   &   & 0 &   &   &   &  \\
  &   &   & 1 &   & -1 &   &   &  \\
  &   & 1 &   &   &   & -1 &   &  \\
\end{matrix}
\end{equation}

We observe that this situation does satisfy the time-symmetric
equation (\ref{eq:timesym}).

Alternatively similar results can be derived by discretizing the
initial velocity $g(x)$ directly using a matching center difference
scheme \cite{Kreyszig99}:

\begin{align}
y^+-y^-=g
\end{align}

and including an arbitrary background field one gets:

\begin{align}\label{eq:initvalkrey}
y^+=\frac{1}{2}(f^<+f^>)+g
\end{align}

where we use the notation $f^+$ and $f^-$ to denote initial
displacement wave contributions aligned with the left-right branch of
the leapfrog-molecule. Observe that (\ref{eq:initvalkrey}) does
satisfy the same integration (\ref{eq:int}) and non-integration
(\ref{eq:nonint}) conditions. Hence a velocity contribution $g$ can be
interpreted as any violation of the rule of the sum of incoming
traveling waves.

This behavior has been observed earlier. Karjalainen observed that an
asymmetric pair of impulses need to be fed into a leapfrog motivated
junction formulations to avoid integration behavior
\cite{Karjalainen03b}. The subsequent physical interpretation is
derived in \cite{KE03} from a center-difference time-discrete velocity
excitation. An interpretation of this result follows next.

\section{Singular Propagation from Integration}


The non-integrating condition can be algorithmically
enforced by a method discovered by Karjalainen and Erkut
\cite{KE03,Karjalainen03}. Hence we will call this the
Karjalainen-Erkut condition. The rule is to present the excitation
through a feed-forward filter \cite{Karjalainen03}:

\begin{align}\label{eq:karer}
H(z)=1-z^{-2}
\end{align}

which can be derived from physical conditions by using a center
difference velocity term \cite{KE03}.

We observe that the Karjalainen-Erkut condition (\ref{eq:karer})
creates two impulses from one and those impulses are center symmetric
and sign-inverted. If we calculate those two impulse responses
separately and then create the sum, we see that the impulsive
propagating solution comes about as the {\em difference of two
Heaviside distributions}. The pulses are represented by their sign
only as the amplitudes are assumed to be matched:

\begin{equation}\label{eq:intconea}
\begin{matrix}
+ &   & + &   & + &   & + &   & +\\
  & + &   & + &   & + &   & + &  \\
  &   & + &   & + &   & + &   &  \\
  &   &   & + &   & + &   &   &  \\
  &   &   &   & + &   &   &   &  \\
\end{matrix}
\end{equation}
\begin{equation*}
\begin{matrix}
  &   &   &   & + &   &   &   &  \\
\end{matrix}
\end{equation*}

\begin{equation}\label{eq:intconeb}
\begin{matrix}
0 &   &- &   &- &   &- &   & 0\\
  & 0 &   &- &   &- &   & 0 &  \\
  &   & 0 &   &- &   & 0 &   &  \\
  &   &   & 0 &   & 0 &   &   &  \\
  &   &   &   & 0 &   &   &   &  \\
\end{matrix}
\end{equation}
\begin{equation*}
\begin{matrix}
  &   &   &   & = &   &   &   &  \\
\end{matrix}
\end{equation*}

\begin{equation}\label{eq:intconeab}
\begin{matrix}
+ &   & 0 &   & 0 &   & 0 &   & +\\
  & + &   & 0 &   & 0 &   & + &  \\
  &   & + &   & 0 &   & + &   &  \\
  &   &   & + &   & + &   &   &  \\
  &   &   &   & + &   &   &   &  \\
\end{matrix}
\end{equation}

Hence we see that in an impulse-response interpretation of the
leap-frog, a Heaviside integration over the characteristic cone of
influence is sensible and the Karjalainen-Erkut condition ensures that
each Heaviside integration is matched with a delayed sign-inverted
response that cancels all interior integration of the first impulse to
leave unaltered the traveling impulse solution.

\section{Effects of the Boundary}

Next the effect of imposing spatial boundary conditions is
studied. For this it is assumed that the solution of the wave equation is
only meaningful and defined for a compact domain $\Omega$. The
length of the domain is denoted by $L=|\Omega|$. For the boundary of the
domain we write $\partial\Omega$ and the interior of the domain is
defined by the quotient $\Omega\setminus\partial\Omega$. On each
distinct point of the boundary $\partial\Omega$ we impose one boundary
condition.  Fixed ends $u(\partial\Omega)=0$ we call Dirichlet
boundary conditions, whereas open ends $u_t(\partial\Omega)=0$ we call
Neumann boundary conditions. Note that a circular domain
$u(|\Omega|)=u(0)$ is a periodic unbounded domain.

The behavior at the boundary can be conveniently studied by extension
of the domain. If the boundary is of Dirichlet type, the value of $u$
needs to vanish at the boundary and hence the extension needs to be
odd. In the case of Neumann conditions $u$ needs to be even. As the
resulting infinite extension is periodic in $2L$ this extension can be
interpreted as a periodic unbounded domain of this length \cite{SS03}.
We will denote the original domain by subscript $0$ and extended
domains by indicies $n\in\mathbb{Z}\setminus 0$. The periodicity
implies $\Omega_m+2=\Omega_m$ for all $m\in\mathbb{Z}$. 

The following discussion will be restricted to the behavior in
response to the velocity term $g$ in (\ref{eq:ivsolve}). Observe that
with periodicity we can write the integral as the sum of contributions
of the periodic domains:

\begin{align}\label{eq:velbound}
\int_{\xi=x-t}^{\eta=x+t}g(s)\,ds=\sum_m\int_{\Omega_m\geq \xi,\eta}g(s)\,ds
\end{align}

Hence we integrate over all contributions above the characteristic
lines from an excitation point of the periodic domain.

With Dirichlet conditions one gets the odd extension \cite{Graff91}:

\begin{align}\label{eq:dirich}
g(x)=-g(2m|\Omega|-x)
\end{align}

and for Neumann conditions we get the even extension:

\begin{align}\label{eq:neumann}
g(x)=g(2m|\Omega|-x)
\end{align}

with $x\in\Omega_0$ and $m\in\mathbb{Z}\setminus 0$.

Integrating up to the point where the characteristic lines have
reached $2|\Omega|$ one gets for Dirichlet boundary conditions:

\begin{align}
u(\Omega)=\int_{\Omega_0} g(s)\,ds-\int_{\frac{1}{2}\Omega_{\pm 1}}g(-s)\,ds=0
\end{align}

Hence integral contributions cancel every $2|\Omega|$ and the maximum amplitude is bounded by the integral
of $g(\cdot)$ over the original domain $\Omega_0$.

The same procedure for Neumann boundary conditions leads to:

\begin{align}
\begin{split}
u(\Omega)&=\int_{\Omega_0} g(s)\,ds+\int_{\frac{1}{2}\Omega_{\pm 1}}g(-s)\,ds\\&=2\int_{\Omega_0} g(s)\,ds
\end{split}
\end{align}

Figure \ref{fig:sumvel} shows the behavior for a string tied at the
ends (Dirichlet conditions) after an initial impulsive distribution.
It reveals many properties of the effect of the boundary on the
integration of velocities under Dirichlet boundary conditions. It
shows the odd-periodic extension of the domain $\Omega_0$ to
$\Omega_n$, $n\in \mathbb{Z}$, it also shows the cancellation and
constructive interference effect of overlapping integration regions.
It also shows the $2|\Omega|$ cancellation of waves. Erkut and
Karjalainen\cite{EK02b} (compare their Figure 7) reported numerical
simulations using the leapfrog molecule with comparable results,
which hence matches the situation of the continuous model.

\subsection{Linear Growth of Displacement}

Observe that the Neumann condition leads to a linear increase in
the displacement as a response to velocity or force data being
present.

The difference between the Dirichlet condition and the Neumann
condition can be interpreted as the difference between an alternating
sum and an accumulative sum.

In the Dirichlet case the sign of the
area integrated over alternates with periodicity $|\Omega|$ and hence
any finite bounded signal $g(\cdot)$ will produce an alternating sum
which is bounded similarly but infinitely periodic.

In the Neumann the signs match and hence the area integrated over
increases with every iteration over the domain by the integral over
the finite bounded signal $g(\cdot)$. Once the support of $g(\cdot)$
has been exhausted, this obviously corresponds to a linear increase.

This is however, not an unphysical situation.  This corresponds to
constant kinetic energy being present in the string and hence implies
that the energy is bound. This can easily be understood as linear
increasing displacement implies constant velocity, which in turn
implies constant energy. Hence energy is conserved\cite{SS03}. It can
be interpreted as a string moving at constant velocity, which is
meaningful as the Neumann conditions imply that the string is not tied
down at the boundaries.  Hence linear buildup in a displacement-like
wave variable is energy-conserving\footnote{Evidently this argument is
valid for any amplitude, also small amplitude oscillations, for which
the wave-equation is valid, as constant displacement does not alter
curvature.}.

Numerically this is still an undesirable situation because even in the
absence of numerical imprecision, the dynamic range of numbers are
bounded and hence an infinite increase cannot be represented.

The case of Neumann boundary conditions is interesting because it
highlights the difference between notions of stability as customary in
discrete signal literature\cite{OS89} and stability in physical
situations. The Neumann displacement response to any bounded input
will be unbounded and hence is evidently not bounded-input
bounded-output (BIBO) stable, see Oppenheim and Schafer\cite{OS89},
p. 20. We suggest that this BIBO-unstable but energy-conserving system
be called {\em physically stable}. The BIBO-instability is a discrete
computational problem and not one of the physical
situation\footnote{Note that the velocity response is in fact
BIBO-stable and hence treatment of the problem in a velocity variable
will not suffer this problem.}.

\section{Implications}

This paper discussed the linear lossless wave equation and its
relationship to discrete models, namely a finite difference scheme
called leapfrog, and the digital waveguide method. It is shown that
the waveguide model corresponds to the finite difference scheme in the
absence of integration. In the continuous case, integration can be expected
to occur when initial velocities or external forces are present.
In this light the observed integrating behavior of finite difference
discretization in the time domain using the leapfrog molecule displays
results consistent with the continuous model. Here we assume that the
wave equation is at least in principle physically meaningful for the
modeled situation. If this is the case one should expect consistent
behavior of the related discrete models.

In relation to this argument, a use of waveguide discretization that
does not include contributions inside the characteristic cone, does
not include the integrating behavior of the model equation. In general
both integrating and non-integrating responses are to be expected and
hence should be present unless they can be explicitly excluded for
physical reasons.

This also implies that the impulse response in just one variable in
general will not carry the full dynamics of the string. Hence any
assumption of the non-integrating impulse response in one variable in
the construction of physical models might contain deviations for it
only covers a reduced set of the solution space.

\begin{acknowledgments}
Much thanks to Sile O'Modhrain for her support and input. The author also has much gratitude to send to Matti Karjalainen and Cumhur Erkut
for stimulating discussions relating to this topic. I am also
grateful for their sending of reprints and the graceful sharing of
novel unpublished manuscripts. This work was made possible by the
kindness of employment of Media Lab Europe and access to its
academic resources.
\end{acknowledgments}

\appendix

\section{Properties of Distributions}\label{sec:distprop}

Let $f$ be a distribution on a real open interval $\Omega$ and let
$\phi$ be in the the set of test functions $\mathcal{D}(\Omega)$ then
one has \cite{Edwards95}:

\begin{align}\label{eq:distint}
\langle\phi,f^\prime\rangle=\int_\Omega f\phi^\prime\,d\mu=-\langle\phi^\prime , f\rangle
\end{align}

and for arbitrary derivatives:

\begin{align}
\langle\phi,\partial^p f\rangle=(-1)^{|p|}\langle\partial^p\phi , f\rangle
\end{align}

the Dirac delta $\delta$ has the property:

\begin{align}
\langle \phi,\delta\rangle = \phi(0)
\end{align}

hence returns the value of $\phi$ at $0$. By the differentiation rule
the higher order derivatives of the Dirac delta returns the higher
order derivatives at $0$ with alternating sign:

\begin{align}\label{eq:distdiffdelta}
\langle \phi,\partial^p\delta\rangle = -(1)^{|p|}\partial^p\phi(0)
\end{align}

Let $H$ be the Heaviside distribution. It is defined as \cite{Hulshof93}:

\begin{align}
\langle \phi,H\rangle = \int_{-\infty}^\infty H(s)\phi(s)ds=\int_0^\infty\phi(s)\,ds
\end{align}

It hence permits the positive part of $\phi$ over the domain. The
derivative of the Heaviside distribution $H$ yields (using (\ref{eq:distint}) and (\ref{eq:distdiffdelta})) the Dirac-delta:

\begin{align}
\langle \phi, H^\prime\rangle = -\langle\phi^\prime,H\rangle=\phi(0)=\langle\phi,\delta\rangle
\end{align}

\section{The Wave Equation and First Order Systems}\label{sec:hidden}

In order to derive the relationship between the wave equation to first
order systems, we discuss two forms of such systems, namely, two
transport equations in one variable and two transport equations in
a mixed pair of variables.

A generic version of a system of
inhomogeneous first order hyperbolic equations reads:

\begin{align}\label{eq:firstorder}
a\partD{y}{x}+b\partD{y}{t}=h_1(x)+h_2(t)\\
c\partD{y}{x}+d\partD{y}{t}=h_3(x)+h_4(t)\label{eq:firstorder2}
\end{align}

For simplicity assume that the force terms are separated in the
independent dimensions. Then a second order version is usually
derived taking the derivative of one equation with respect to $t$ and
the other one with respect to $x$. The cross-term $y_{xt}$ can be
eliminated and one gets two equations:

\begin{align}\label{eq:secondorder}
\frac{b}{d}\partN{y}{t}{2}-\frac{c}{a}\partN{y}{x}{2}=\frac{1}{d}\partop{t}h_2(t)-\frac{1}{a}\partop{x}h_3(x)\\
-\frac{a}{c}\partN{y}{t}{2}+\frac{d}{b}\partN{y}{x}{2}=\frac{1}{c}\partop{x}h_1(x)-\frac{1}{b}\partop{t}h_4(t)\label{eq:secondorder2}
\end{align}

The key observation is that one second order equation
(\ref{eq:secondorder}) or (\ref{eq:secondorder2}) is not strictly
equal to the system of first order equations (\ref{eq:firstorder}) and
(\ref{eq:firstorder2}). It is only equal up to two functions
(whichever got eliminated, $h_1,h_4$ or $h_2,h_3$). They are
equivalent up to two "constants of integration".

For systems of first order linear equations in two independent
variables a related proof holds. An intuitive interpretation is that
in fact for first order equations of the type:

\begin{align}\label{eq:coup1}
u_x+w_t&=g_1(t)\\
w_x+u_t&=g_2(x)\label{eq:coup2}
\end{align}

one sees that the reduction to second order equations in $u$
by differentiating (\ref{eq:coup1}) with respect to $x$ and
(\ref{eq:coup2}) with respect to $t$ one gets:

\begin{align}
u_{xx}-u_{tt}&=0\\
-w_{xx}+w_{tt}&=\dot{g_1}(t)-g_2^\prime(x)
\end{align}

Note that differentiation eliminated $g_1$ and $g_2$ in one case and
hence the homogeneous wave equation is again indistinguishable for
both the homogeneous and a class of inhomogeneous systems of first
order equations and in this sense they are equivalent only up to two
functions.


\newpage

\newpage
\begin{figure}
\centering
\includegraphics[width=1.25in]{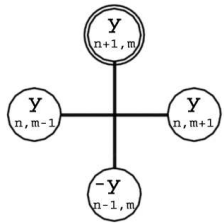}
\caption{Leapfrog computational molecule for the one-dimensional wave equation.}
\label{fig:nonprop}
\end{figure}

\newpage
\begin{figure}
\centering
\includegraphics[width=3.375in]{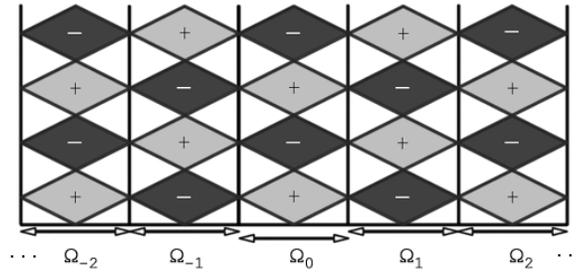}
\caption{Sum of velocity domains. $\Omega_0$ is the original string domain and $\Omega_n$ with $n\in \mathbb{Z}\setminus 0$ are domains created by continuation of the domain obeying the boundary condition $u(\partial\Omega)=0$.}
\label{fig:sumvel}
\end{figure}

\end{document}